\documentclass{PoS}

\title{Characterizing X-ray Variability of TeV Blazars}

\ShortTitle{X-ray Variability of TeV Blazars}

\author{\speaker{Jun Kataoka}\\
        Tokyo Institute of Technology, 2-12-1, Ohokayama, Meguro, 
        Tokyo, Japan\\
        E-mail: \email{kataoka@phys.titech.ac.jp}}

\abstract{In this review, I will discuss how to characterize synchrotron 
X-ray variability of TeV blazars by using the observed/simulated light 
curves. Apparently, temporal studies provide independent and 
complementary information to the spectral studies, 
but surprisingly little attention has been paid especially for the 
blazar study. Only exception is a classical argument for presence 
of ``time lag'', which may (or may not) reflect the 
diffrence of synchrotron cooling timescale. 
Also very recently, it was suggested that the 
X-ray variability of TeV blazars indicates a strong red-noise, compared 
to a fractal, flickering-noise of Seyfert galaxies. 
Various temporal techniques are proposed in 
literature, e.g., the power spectrum density (PSD), the
structure function (SF),  and the discrete correlation function (DCF) 
and other analysis tools, but special care must be taken if the 
data are not well sampled and observation is relatively short compared
to a characteristic timescale of the system. 
Also, the situation is being more complicated for low-Earth orbit 
satellites, e.g., $ASCA$, $RXTE$ and $BeppoSAX$, since the light curve 
inevitably contains ``periodic gap'' 
due to the Earth occultation (every $\simeq$ 6ksec). 
I will present detailed approaches to see how the "gap" 
and the "finite length" of the data affects the results of temporal analysis, 
and to what extent we can believe in our results.  Finally, 
I will briefly comment on the high-sensitivity X-ray observations 
with $MAXI$, that may shed new light on the forthcoming $GLAST$ era.}

\FullConference{Workshop on Blazar Variability across the Electromagnetic Spectrum\\
		 April 22-25 2008\\
		 Palaiseau, France}

\begin{document}

\section{Introduction}

Blazars are commonly variable from radio to $\gamma$-rays.
The variability timescale is shortened and the radiation is strongly
enhanced by relativistic beaming. For extragalactic
TeV sources, the X-ray/TeV $\gamma$-ray bands correspond to the highest
energy ends of the synchrotron/inverse-Compton emission,
which are produced by electrons accelerated up to the maximum energy
(e.g., Inoue \& Takahara 1996; Krik, Rieger \& Mastichiadis 1998). 
At the highest energy ends, variability is expected to be most 
pronounced, and in fact, such large flux variations are observed, 
on a timescale of hours to days (e.g., Kataoka et al. 2001; 
Tanihata et al. 2001) or even shorter (minutes scale; 
Aharonian et al. 2007; Albert et al. 2007). Thus the X-ray/TeV 
variability can be the most direct way to probe the dynamics operating 
in jet plasma, in particular compact regions of shock acceleration 
which are presumably close to the central engine.

`Snapshot' multiwavelength spectra principally provide us with clues
on the emission mechanisms and physical parameters inside relativistic jets.
On the other hand, detailed studies of time variability
not only lead to complementary information for the objectives above,
but should also offer us
a more direct window on the physical processes operating in the jet
as well as on the dynamics the jet itself.
However, short time-coverage and under-sampling have
prevented detailed temporal studies of blazars.
Only a few such studies have been made in the past for blazars,
e.g., evaluation of the energy dependent
``time-lags'' based on the synchrotron cooling picture. 
For example, by using $ASCA$ data,  Takahashi et al. (1996) argued 
the soft X-ray ($<$ 1 keV) variation of Mrk~421, observed to lag 
behind that of the hard X-rays ($\ge$ 2 keV) by $\sim$ 4 ks, that 
may well be ascribed to the energy dependence of the 
synchrotron cooling timescale. More recently,
Kataoka et al. (2000) interpreted an observed soft-lag and spectral
evolution of PKS~2155-304 by a newly developed time-dependent
synchrotron self-Compton (SSC) model.

The above $paradigm$ of ``soft-lag'' was concerned, however,
by several aspects. First, intensive X-ray monitoring of
blazars has revealed not only soft lags
but in some cases hard lags (Takahashi et al. 2000) 
which may be a manifestation of another process, 
e.g., energy dependent acceleration. Very recently, signature of hard
lag was clearly observed in 1ES 1218+304, but this is so far 
an only example of manifestation of possible acceleration timescale in any TeV 
blazars (Sato et al. 2008; also in this volume).
Second, as Edelson et al. (2001) voiced concerns, there was a 
question about the reliability of lags that are smaller than the 
orbital periods ($\sim$ 6 ks) of low Earth orbit satellites.
This was refuted by Tanihata et al. (2001) and Zhang et al. (2004) who
showed that, although periodic gaps introduce larger uncertainties than evenly
sampled data, lags on hour-scale cannot be the result of
periodic gaps. A time resolved cross
correlation analysis of uninterrupted Mrk~421 data obtained by $XMM$-$Newton$
revealed lags of both signs, changing
on timescales of up to a few 10$^3$ s (Brinkmann et al. 2005).
Hence the situation is very complex and still under debate.

Variability studies covering larger dynamic range and broader 
span of timescales have become common for Seyfert galaxies and 
Galactic black-holes (Edelson \& Nandra\ 1999; Markowitz  et al.\
2004; McHardy et al. 2005; 2008 in this volume). 
From power spectrum density 
(PSD) analyses, it is well known that rapid fluctuations with frequency 
dependences $P(f)$ $\propto$ $f^{ -1 \sim -2}$, 
are characteristic of time variability in 
accreting black hole systems (e.g., Hayashida et
al. 1998). Although their physical origin is still under debate, some
 tentative scenarios have been suggested to account for these generic, 
fractal features (e.g., Kawaguchi et al.\ 2000). Similar studies have 
also been proposed for blazars, but still underway. It has been suggested 
that X-ray variability of TeV blazars indicates a strong red-noise 
($P(f)$ $\propto$ $f^ {-2 \sim -3}$) behavior, 
compared to a fractal, 
flickering-noise of Seyfert galaxies (Kataoka et al. 2001).

These temporal studies are obviously important, however, special 
care must be taken if the data are not well sampled and relatively 
short compared to the variability timescale of the system. 
The prime motivation of this talk is to delineate the characteristic 
X-ray variability of TeV blazars, using a simple Monte Carlro 
simulation to evaluate the possible effects caused by observing 
time windows. Fortunately, we have now the $GLAST$ mission successfully 
launched in June 2008, as well as various excellent missions/telescopes 
available through radio to TeV energy bands. Moreover, future X-ray 
missions including Monitor of All-sky X-ray Image ($MAXI$) is ready for 
launch early next year. A great advantage of $GLAST$ and $MAXI$ is to 
provide very uniform exposure all over the sky, that may shed new 
light on the temporal studies of blazars especially on longer timescale 
from a month to years.

\section{Analysis Tools}

\subsection{Power Spectrum Density (PSD)}

Power Spectrum Density (PSD) analysis is the most common technique used
to characterize the variability of the system. 
An important issue is the data gaps, which are
unavoidable for low-orbit X-ray satellites, such as $ASCA$, $RXTE$, 
and $BeppoSAX$. In these low-orbit satellites, Earth occultation makes 
periodic gaps every $\simeq$ 6 ksec, even if we hope to make 
$continuous$ monitoring observations.  
Similarly, previous long-look observations of 
various TeV blazars (e.g., Mrk 501 and PKS~2155$-$304 in Kataoka 
et al. 2001) inevitably faced serious artificial gaps, since the
observations are spaced typically 3 or 4 orbits apart.
To reduce the effects caused by such windowing, it is recommended to use
a technique for calculating the PSD of unevenly sampled 
light curves.

Following  Hayashida et al.\ (1998), the NPSD ($Normalized$ Power Spectrum
Density) at frequency $f$ is defined as
\begin{eqnarray}
P(f) = \frac{[a^2(f)+b^2(f)-\sigma^2_{\rm stat}/n]T}{F_{\rm av}^2},\nonumber \\
a(f) = \frac{1}{n}\sum_{j=0}^{n-1} F_j {\rm cos} (2\pi f t_j), \nonumber \\
b(f) = \frac{1}{n}\sum_{j=0}^{n-1} F_j {\rm sin} (2\pi f t_j), \nonumber \\
\end{eqnarray}
where $F_j$ is the source count rate at time $t_j$ (0$\le$$j$$\le$$n$$-$1),
$T$ is the data length of the time series and  $F_{\rm av}$ is the mean
value of the source counting rate. The power due to the photon counting 
statistics is given by $\sigma_{\rm stat}^2$.
With our definition, integration of power over the positive frequencies is
equal to half of the light curve excess variance (e.g., Nandra et al.\ 1997).

To calculate the NPSD of certain data sets, it is recommended to make 
light curves of two different bin sizes shorter/longer than orbital 
gaps (e.g., 256 and 5760 sec, respectively). 
Each light curve is divided into ``segments'', which
are defined as the continuous part of the light curve. 
One can then calculate the power at frequencies $f$ = $k$/$T$ 
(1 $\le$ $k$ $\le$ $n$/2) for each segment and take the average.
In this manner, the light curve binned at shorter timescale
 is divided into different segments every 5760~sec, 
corresponding to the gap due to orbital period.  
On the other hand, the light curve binned at 5760 sec is smoothly 
connected up to the total observation length $T$, if further 
artificial gaps are not involved. This technique
produces a large blank in the NPSD at around 2$\times$10$^{-4}$~Hz
(the inverse of the orbital period),
but the effects caused by the sampling window are minimized.

Figure~1($left$) shows examples of $observed$ X-ray 
light curves of three TeV blazars, obtained during long 
($\sim$ 10 day) monitoring with $ASCA$ in 1998$-$2000 
($red$: Mrk 421, $blue$: Mrk 501; $green$: PKS 2155$-$304, respectively).
Corresponding NPSDs are given in Figure~1($right$). 
The upper frequency limit is the Nyquist frequency (2$\times$10$^{-3}$~Hz 
for 256~sec bins) and the lower frequency is about half the inverse of the 
longest continuous segments. These NPSD are binned in logarithmic intervals 
of 0.2 (i.e.\ factors of 1.6) to reduce the noise. Note the NPSDs 
follow a power-law that decreases with increasing frequency 
in the high-frequency range 
(typically, $P(f)$ $\propto$ $f^{ -2.5}$). Possible signs of a 
roll-over can be seen at the low-frequency end ($f_{\rm  br}$ 
$\sim$ 10$^{-5}$~Hz). Below this break, the NPSD becomes significantly 
flatter, such that $P(f)$ $\propto$ $f^{  -1.0}$.
Since all the NPSDs have very steep power-law slopes, only little 
power exists above 10$^{-3}$ Hz. This is very different from 
the PSDs of Seyfert galaxies, for which powers are well above 
the counting noise up to 10$^{-2}$~Hz (e.g., Hayashida et al.\ 1998; 
Nowak \& Chiang 2000). 

\begin{figure}
  \includegraphics[height=.33\textheight]{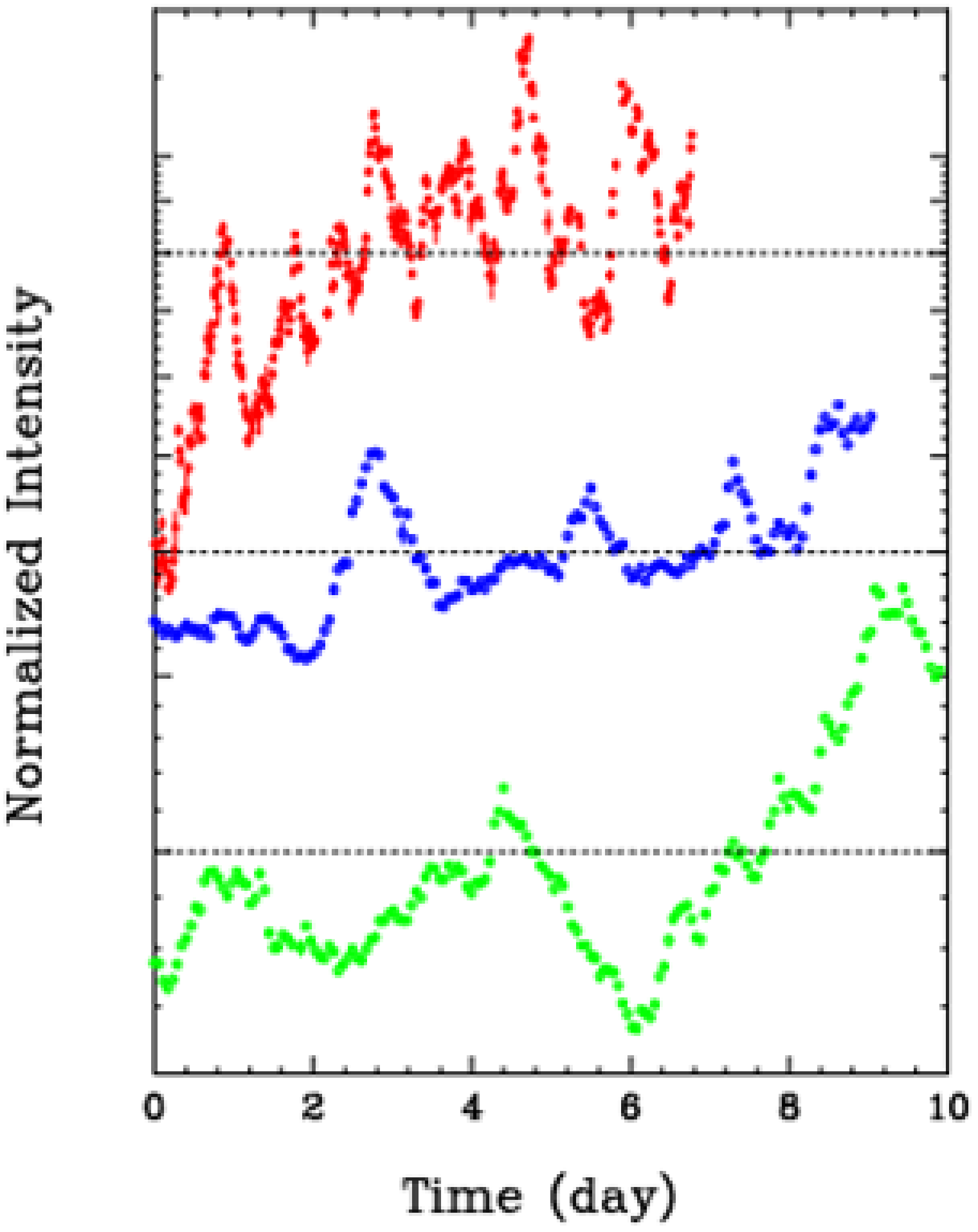}
  \includegraphics[height=.33\textheight]{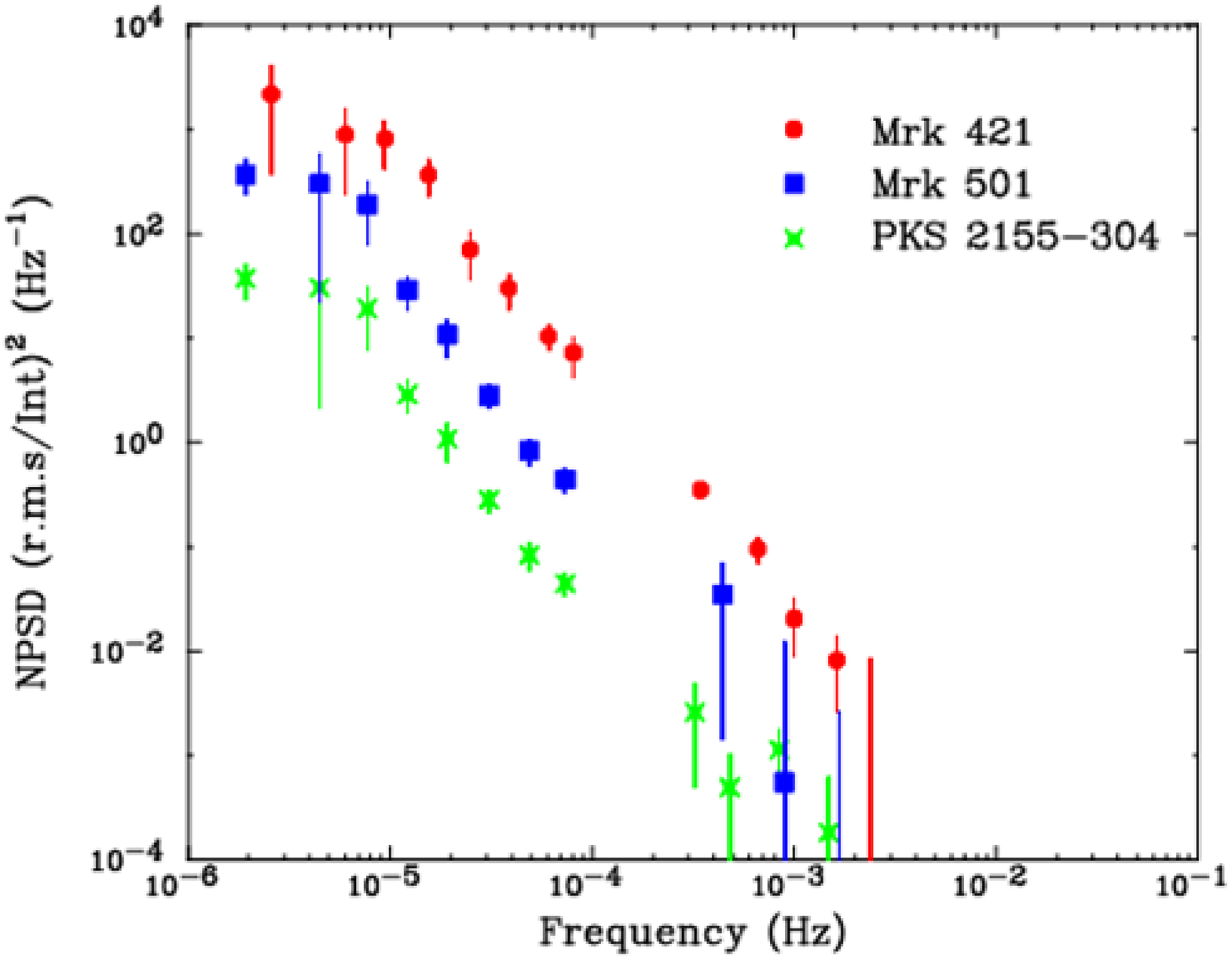}
  \caption{$left$: X-ray light curve  of Mrk 421 ($red$), Mrk 501 ($blue$), 
and PKS 2155$-$304 ($green$) during intensive monitoring campaign with $ASCA$, 
measured in 0.7$-$7.5 keV band. Flux is normalized by their 
average intensities (dashed lines).
$right$: Normalized PSD calculated from the light curves in the 
left panel. See Kataoka et al. 2001 and Tanihata 2001 for more details.}
\end{figure}

Finally, we revisit the effects caused by sampling windows. As 
mentioned above, our PSD technique is less affected by the sampling 
windows, because only the continuous parts of the light curve are used for the
calculation. In fact, this seems to have negligible effects for the
present data, because the interruptions are almost even and the 
observing efficiency is high ($\sim$ 0.5). The most rigorous estimate 
of this effect would be obtained by simulating the light curves 
characterized with a certain PSD, filtered by the same window 
as the actual observation. The resulting PSDs could then be 
compared with that we assumed.  For this purpose, using a Monte 
Carlo technique, we generate a set of random numbers uniformly 
distributed between 0 and 2$\pi$ and  use them as the random 
phases of the Fourier components. A fake light curve is then 
generated by a Fourier transformation, with the constraint that 
the power in each frequency bin decreases as specified by the PSD. 
We simply choose a deterministic amplitude for each frequency 
and randomize only the phases, a common approach 
(e.g., Done et al. 1989). It may be most rigorous to also assume 
``random amplitudes'' distributed within 1~$\sigma$ of the input PSD 
(Timmer \& K\H{o}nig 1995), but simulations based on their algorithm
remain as a future work.

Figure~2 ($left$) shows an example light curve thus produced,
with a hypothesized PSD of the form;  
$P(f) = P_0 f^{-2.5}$ ($f$ $\ge$ $f_{\rm br}$) 
and $P(f) = P_0$ ($f$ $\le$ $f_{\rm br}$), where 
$f_{\rm br}$ = 10$^{-5}$ Hz.  We have made hundreds of 
such pseudo light curves and calculated the PSD as actual 
observational data. Right panel shows the PSDs plotted for 10 sets of 
such light curves. One can see that even when the orbital gaps are 
present, the resultant PSD is not affected seriously, and hence we 
can safely determine the original PSD which produces the 
observed light curves.   

\begin{figure}
  \includegraphics[height=.293\textheight]{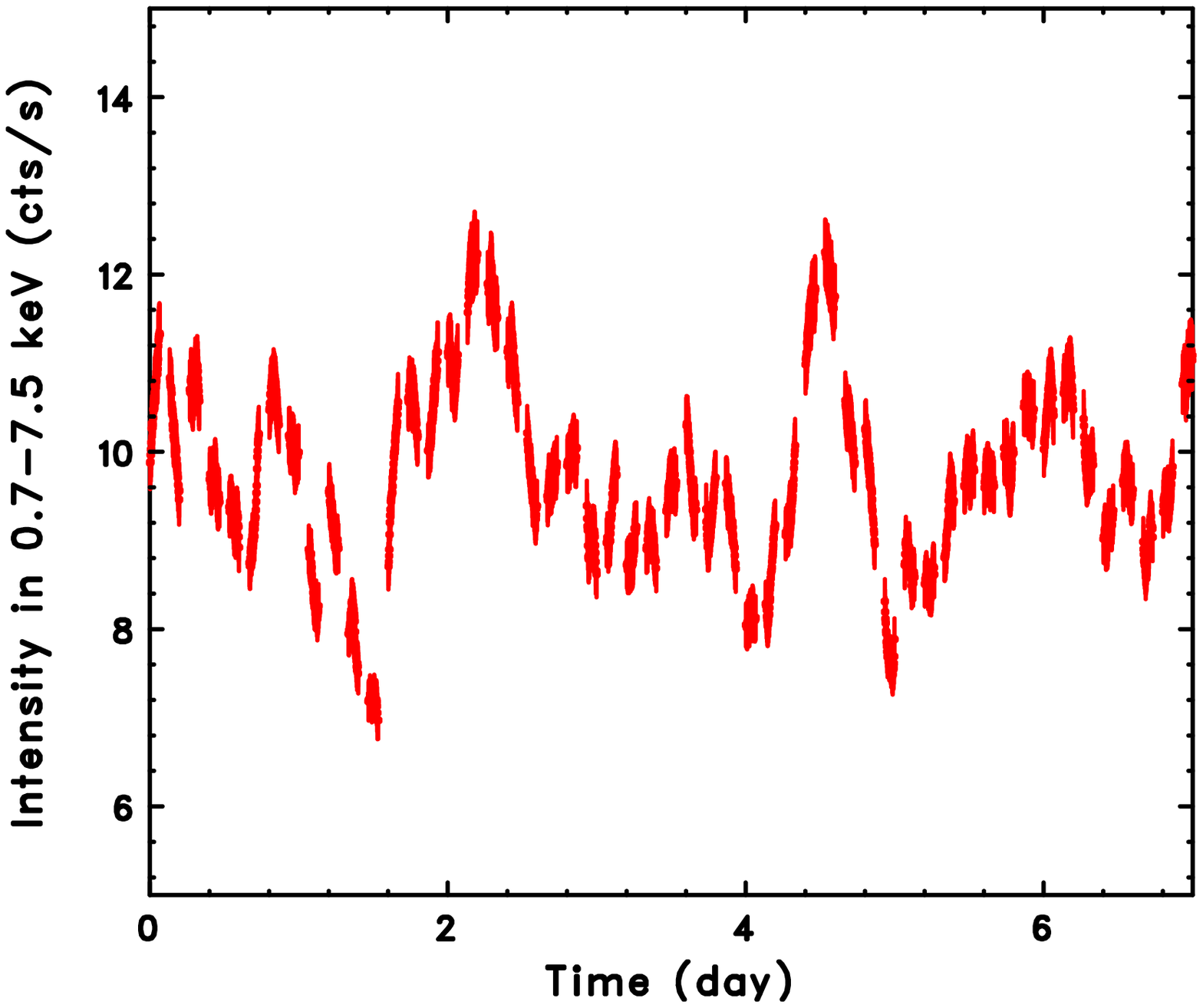}
  \includegraphics[height=.293\textheight]{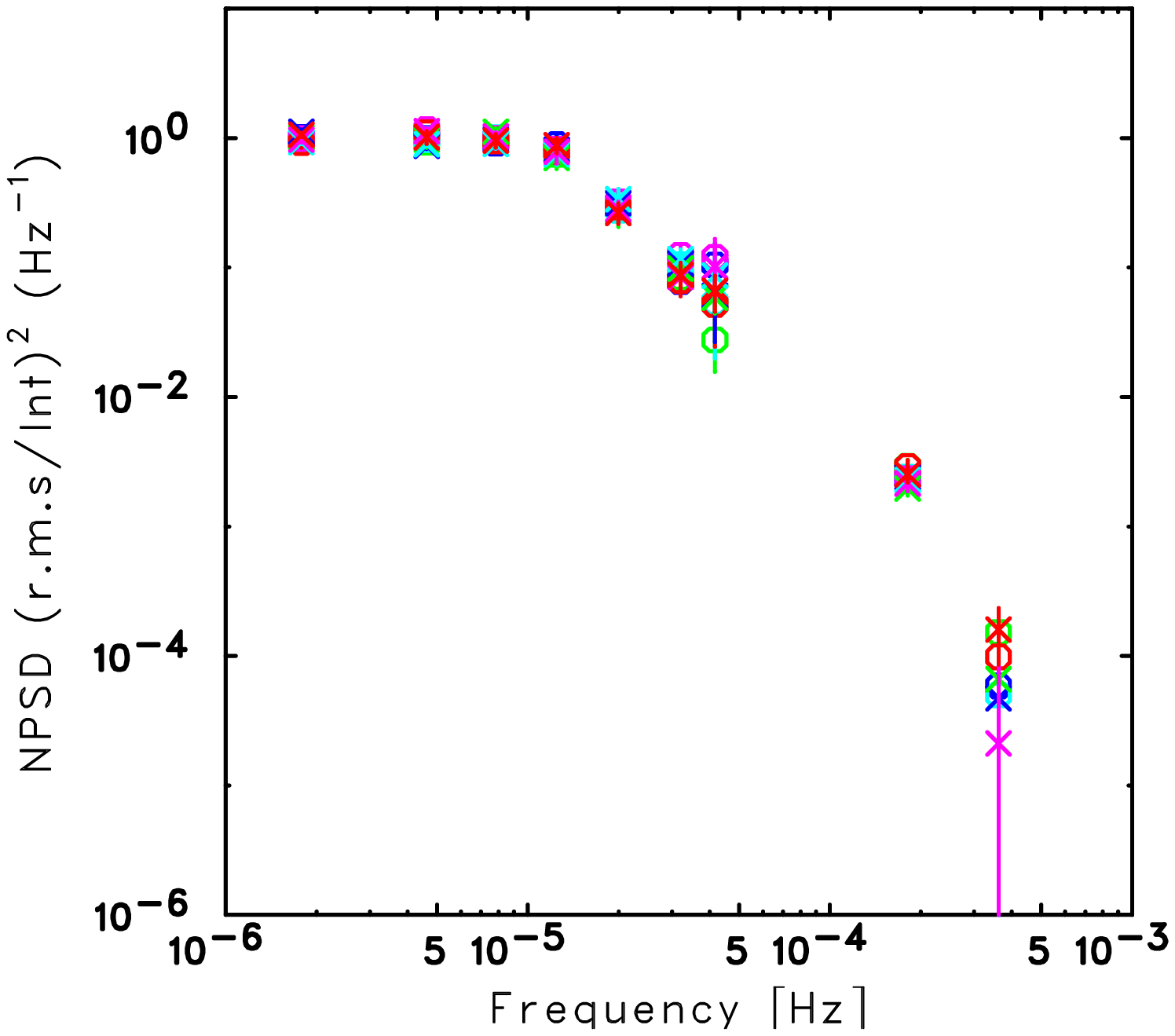}
  \caption{$left$: An example of simulated light curve described 
in the text. $right$: Examples of normalized PSD calculated 
from 10 simulated light curves.}
\end{figure}

\subsection{Structure Function (SF)}

Next we examine the use of a numerical technique called 
the structure function (hereafter, SF). The SF can potentially
 provide information on the nature of the physical process causing 
any observed variability. While in theory the SF is completely 
equivalent to traditional Fourier analysis methods (e.g., the NPSD;
$\S$~2.1), it has several significant advantages. Firstly, it is much 
easier to calculate. Secondly, the SF is less affected by gaps
in the light curves (e.g., Hughes et al.\ 1992). The definitions of 
SFs and their properties are given by Simonetti et al.\ (1985). 
The first order SF is defined as
\begin{equation}
{\rm SF}(\tau) = \frac{1}{N}\sum[a(t) - a(t+\tau)]^2,
\label{equation:1-1}
\end{equation}
where $a(t)$ is a point of the time series (light curves) $\{$$a$$\}$ and
the summation is made over all pairs separated in time by $\tau$.
$N$ is the number of such pairs. More rigorously, minor 
modification is suggested in 
Tanihata et al. (2001), that one should use a continuous weighting 
factor proportional to its significance of each data point when the 
flux uncertainties are non-uniform. Note that the SF is free from the DC
component in the time series, whereas techniques such as the auto-correlation
function (ACF) and the PSD are not. 

The SF is closely related with the power spectrum density (PSD) distribution.
If the structure function has a power-law form, SF($\tau$) $\propto$
$\tau^{ \beta}$ ($\beta$ $>$ 0), then the  power spectrum has the distribution
$P(f)$ $\propto$ $f^{ -\alpha}$, where $f$ is frequency and $\alpha$
$\simeq$ $\beta$ + 1.  We note that this approximation is invalid
when $\alpha$ is smaller than 1. In fact, both the SF and the NPSD should
have zero slope for white noise, because it has zero correlation timescale.
However, the relation holds within an error of $\Delta\alpha$ $\simeq$
0.2 when $\alpha$ is larger than $\sim$1.5 (e.g., Paltani et al.\ 1997;
Cagnoni, Papadakis \& Fruscione 2001; Iyomoto \& Makishima 2000).
Therefore the SF gives a crude but convenient estimate of the corresponding
PSD distribution which characterizes the variability.

In general, the SF gradually changes its slope ($\beta$) with time interval
$\tau$. On the shortest timescale, variability can be well approximated
by a linear function of time; $a(t)$ $\propto$ $t$.
In this time domain, the resulting SF is $\propto$ $\tau^2$, which is the
steepest portion in the SF curve. For longer timescales,
the slope of the SF becomes flatter ($\beta$\,$<$\,2) reflecting the 
physical process operating in the system. When $\tau$ exceeds the 
longest time variability of the system, the SF further flattens, 
with $\beta$\,$\sim$\,0, which is the flattest portion in the SF curve 
(white noise). At this end, the amplitude of the SF is equal to twice 
the variance of the fluctuation. In Figure~3 ($left$), the SF is calculated 
 for the light curves of Mrk 421 presented in Figure~1 ($left$). 
The resulting SF is normalized by the square of the mean fluxes, 
and are binned at logarithmically equal intervals. Note that the SF is 
characterized with a steep increase ($\beta$ $>$ 1) in the 
time region of 10$^{-2}$ $<$ $\tau$/day $<$ 1, roughly consistent 
with the corresponding NPSDs given in Figure~2 
($P(f)$ $\propto$ $f^{  -2.5}$).

\begin{figure}
  \includegraphics[height=.274\textheight]{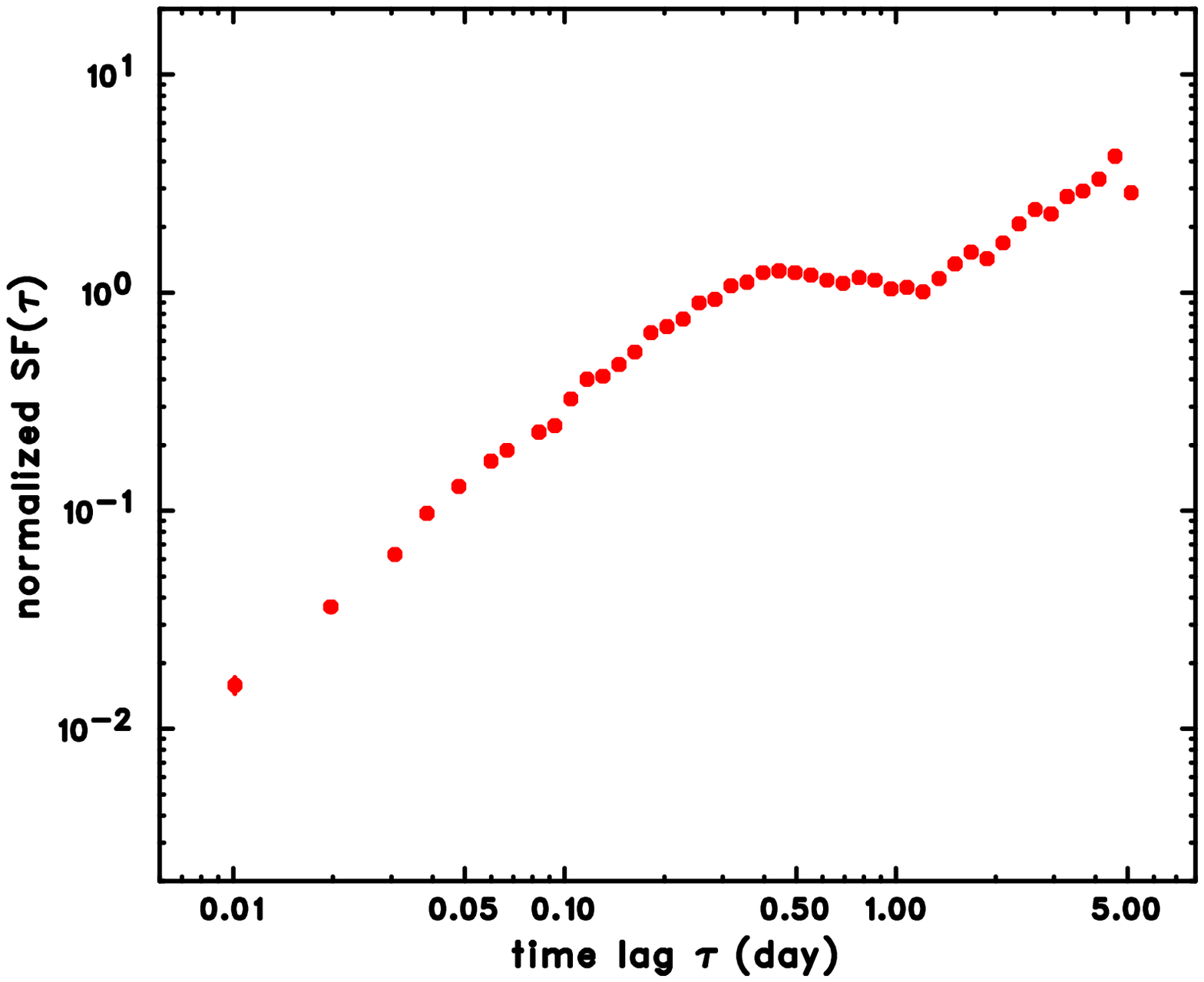}
  \includegraphics[height=.274\textheight]{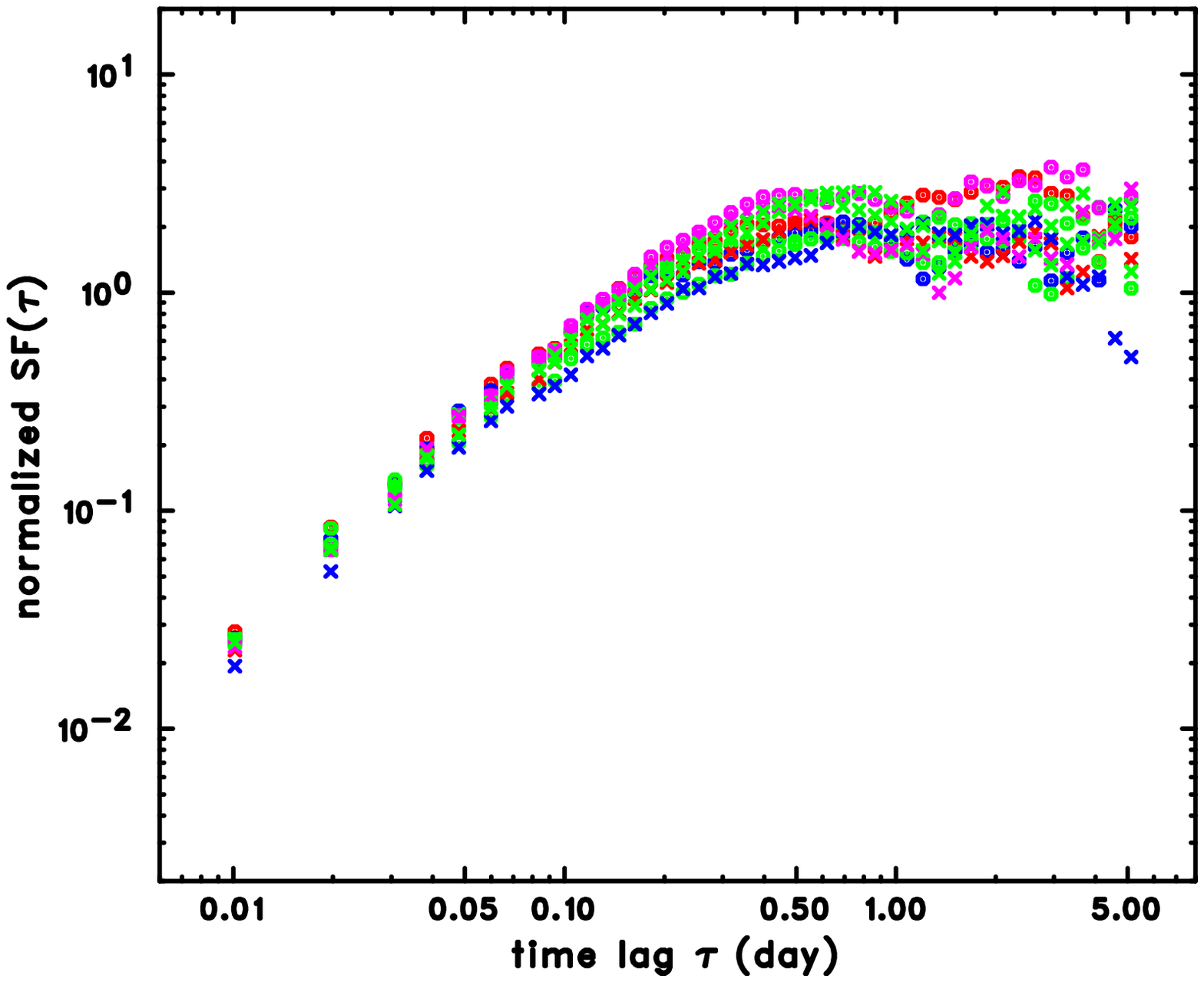}
  \caption{$left$: Structure function calculated from the observed
  light curve of Mrk 421 with $ASCA$ (see Figure 1 $left$). 
$right$: Structure function calculated from the simulated light curves 
to see the effects caused by finite length of data.}
\end{figure}

The SFs of the X-ray light curves show a variety of features. 
For example, the SF of Mrk~421 (Figure 3 ($left$)) 
shows a complex SF that cannot even 
be described as a simple power-law, as it flattens around
0.5 day, then steepens again around 2 days.  A similar ``roll-over'' 
can be seen for the SF of Mrk~501 and PKS 2155$-$304 around 1~day 
(Tanihata et al. 2001). Importantly, these turn-overs reflect the 
typical timescale of repeated flares, corresponding to the break 
in the NPSDs described in $\S$~2.1. The complicated features 
(rapid rise and decay) at large $\tau$ may not be real and may 
result from the insufficiently long sampling of data. 
The number of pairs in Equation (2.2)
decreases with increasing $\tau$, and hence the resulting SF becomes 
uncertain as $\tau$ approaches $T$, where $T$ is the total 
length of time series.
The statistical significance of these features can be easily tested 
using the Monte Carlo simulation. Figure 3 ($right$) shows a set of SFs
calculated by assuming the same PSD described in the previous section 
($P(f) = P_0 f^{ -2.5}$ for $f$ $\ge$ $f_{\rm br}$, where $f_{\rm br}$ 
= 10$^{-5}$ Hz). Note we have simulated the light curves more than 5 
times longer than actual observation. 
One can see although the resultant SFs well 
agree below the break, wide variety exists if $\tau$ $\gtrsim$ 1/3 $T$ 
due to uncertainties caused by  finite length of data. Therefore 
special care must be taken for possible artifacts near 
$\tau_{\rm max }$ = $T$.

We next calculate the structure functions using all available X-ray 
data set between 1993 and 1998 for Mrk 421. Using 5 year's 
$ASCA$ data, we can investigate the variability in the widest 
time domain over more than  five orders;  $10^{-2}$ $\le$ $\tau$/day 
$\le$ 10$^3$. Figure 4 ($left$) shows a light curve thus produced, while 
the SF is given in Figure 4 ($right$). $Filled$ $circles$ are observational 
data, normalized by the square of the mean fluxes, and are binned at 
logarithmically equal intervals. The SF shows a rapid increase up to 
$\tau$/day $\simeq$ 1, then gradually flatten to the observed longest 
timescale of $\tau$/day\,$\ge$\,1000. Fluctuations at large 
$\tau$ ($\tau$/day\,$\ge$\,10) are due to the extremely sparse 
sampling of data. Although we cannot apply the usual PSD 
technique to such under-sampled data, it appears the SF still 
can be a viable estimator.

In order to demonstrate the uncertainties caused by such sparse
sampling, and to firmly establish the reality of the ``roll-over'', 
we simulate the long-term light curves following the 
Monte Calro method described above. We first applied this 
technique assuming a PSD of the form $P(f)$ $\propto$ $f^{ -\alpha}$, 
where $\alpha$ is determined from 
the best fit NPSD parameters given in Figure~1.  Based on a set 
of a thousand fake light curves, we computed the expected mean value, 
$<$$SF_{\rm sim}(\tau)$$>$, and variance, $\sigma$$_{SF(\tau)}$, 
of all the simulated SFs at each $\tau$. The results are 
superimposed in Figure~5(b) as $crosses$. Errors on simulated 
data points are equal to $\pm$$\sigma$$_{SF(\tau)}$. 
One finds that errors become larger at large $\tau$, meaning 
that the SF tends to involve fake bumps and wiggles near the 
longest observed timescale. Large deviations
between the actual SFs ($filled$ $circles$) and the simulated ones 
($crosses$) are apparent, but quantitative comparison with actual 
data is necessary.

\begin{figure}
  \includegraphics[height=.28\textheight]{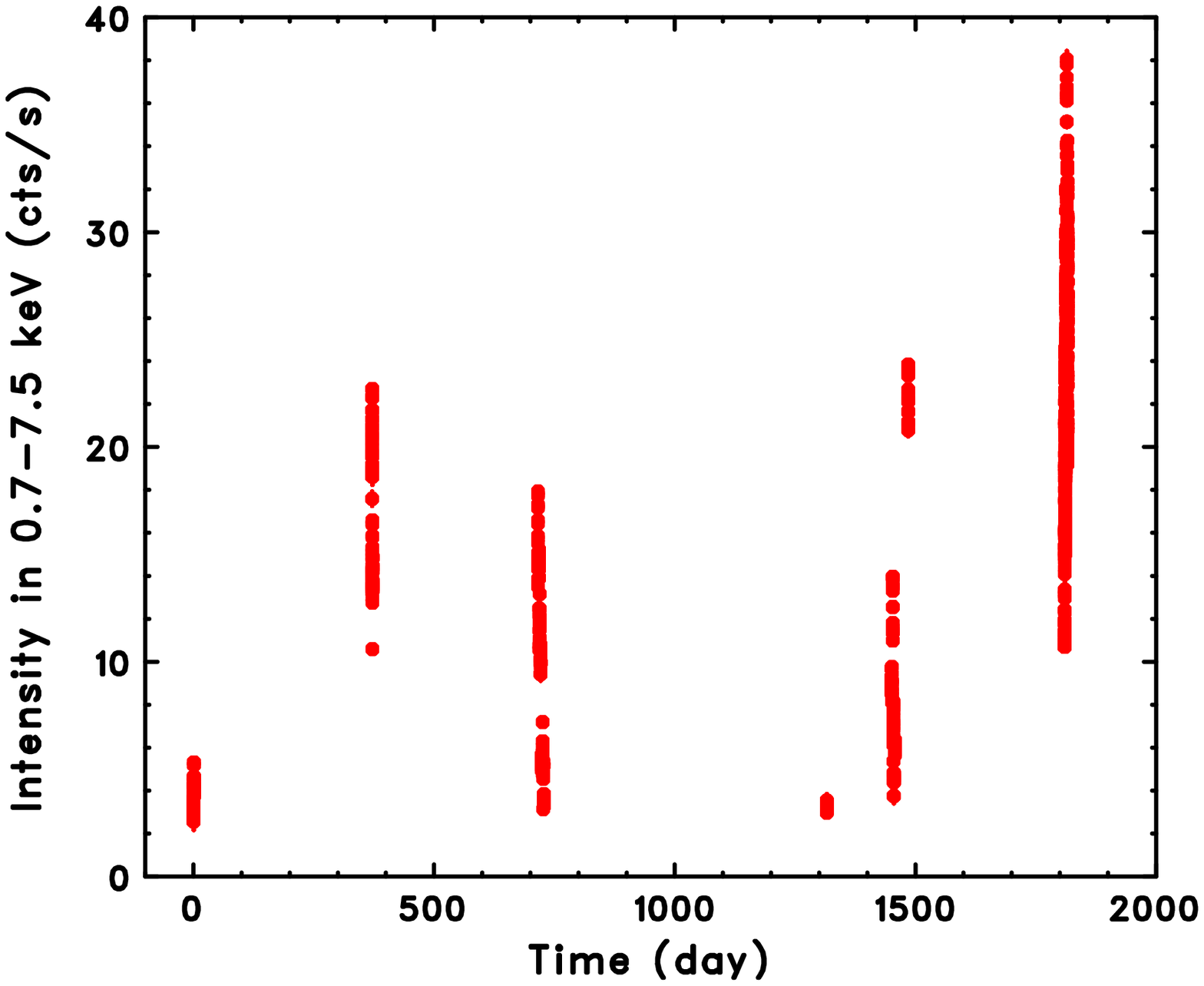}
  \includegraphics[height=.29\textheight]{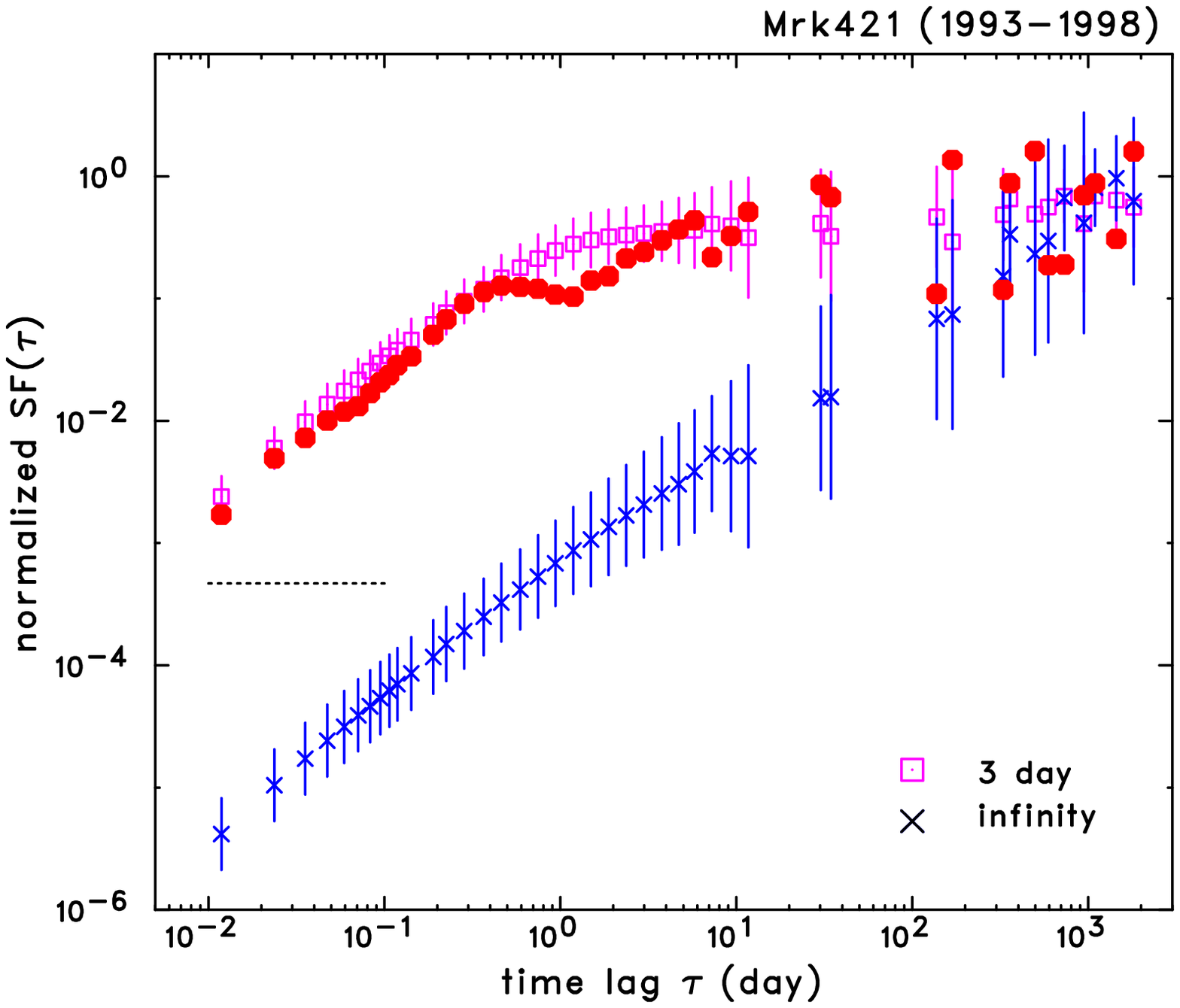}
  \caption{$left$: Long-term X-ray flux variation of Mrk 421 between 1993 and
 1998, measured in 0.7$-$7.5 keV band. 
$right$: Structure function of Mrk 421 based on long-term light curves
 presented in the $left$ panel. $Filled$ $circles$ represent the
 observational data, $crosses$ represent simulated SF assuming a
 single-power-law NPSD, and $open$ $squares$ represent simulated SF assuming
 a broken power-law NPSD. Full details are given in the text.}
\end{figure}

To evaluate the statistical significance of the goodness of fit, and
to test the reality of complicated features in the SF, we then calculate
the sum of squared differences,\\
$\chi_{\rm  sim}^2$ = $\sum_{k}$$\{$log[$<$$SF_{\rm sim}$($\tau_{k}$)$>$]$-$log[$SF$($\tau_{k}$)]$\}$$^2$.
Strictly speaking, ``$\chi_{\rm sim}^2$'' defined here is different from
the traditional $\chi^2$, but the statistical
meaning is the same. For the actual SFs, these values are
$\chi_{\rm sim}^2$ = 1608 for Mrk~421. We then generated 
$another$ set of 1,000 simulated light curves and hence fake SFs 
to evaluate the the distribution of $\chi_{\rm sim}^2$ values. From this
simulation, the probability that the X-ray light curves are the realization
of the assumed PSDs (i.e., a simple power-law) 
is $P(\chi^2)$ $<$ 10$^{-3}$. 
We thus introduce a ``break'', below which  the slope of the 
PSD becomes flatter. Since the exact position of 
a break is not well constrained, we simulate various cases of 
$f_{\rm br}$ = 3.9$\times$10$^{-5}$, 1.2$\times$10$^{-5}$, and  
3.9$\times$10$^{-6}$~Hz, which correspond to the break in the 
SF at $\tau$/day $\simeq$ 0.3, 1, 3, respectively.
As a result, the statistical significance is significantly improved.
Result is given in Figure~4 ($right$) as $open$ $squares$.
For Mrk~421, best fit $\chi^2$ was obtained when $f_{\rm br}$
= 3.9$\times$10$^{-6}$ Hz ($\chi^2$ = 47; $P(\chi^2)$ = 0.59). 
We thus conclude that (1) the PSD of the TeV sources
have at least one roll-over at 10$^{-6}$ Hz $\le$ $f_{\rm br}$ $\le$
10$^{-5}$ Hz (1 $\le$ $\tau$/day $\le$ 10),  and (2)
 the PSD changes its slope from $\propto$ $f^{-1 \sim -2}$ 
($f$ $<$ $f_{\rm br}$) to $\propto$
$f^{-2 \sim -3}$ ($f$ $>$ $f_{\rm br}$) around the roll-over.

\subsection{Discrete Correlation Function (DCF)}

In order to compare the time series in various energy bands quantitatively,
we introduce the discrete correlation function given by Edelson \& 
Krolik (1988). This technique was specifically designed to analyze 
unevenly sampled data sets. The first step is to calculate the set 
of unbinned discrete correlations (UDCF) between each data point 
in the two data streams. This is defined in the time domain as
\begin{equation}
{\rm UDCF}_{ij} = \frac{(a_i - \bar{a})(b_j - \bar{b})}{\sqrt{\sigma_a^2 \sigma_b^2}},
\end{equation}
where $a_i$ and $b_j$ are points of the data set $\{$$a$$\}$ and $\{$$b$$\}$,
$\bar{a}$ and $\bar{b}$ are the means of the data sets,
and $\sigma_a$ and $\sigma_b$  are the standard deviation of each data set.
The discrete correlation function (DCF) for each time lag $\tau$ is defined
as an average of  the UDCF that have the same $\tau$,
\begin{equation}
{\rm DCF}(\tau) = \frac{1}{M}\sum {\rm UDCF}_{ij}(\tau),
\end{equation}
where $M$ is the number of pairs in the bin.

The DCF advantages are that it uses all the data points available, does not
introduce new errors through interpolation, and calculates a meaningful error
estimates. The standard error for each bin is calculated as
\begin{equation}
\sigma_{\rm DCF} = \frac{1}{M-1}(\sum [{\rm UDCF}_{ij} - {\rm DCF}(\tau)]^2)^{1/2}.
\end{equation}

\begin{figure}
  \includegraphics[height=.305\textheight]{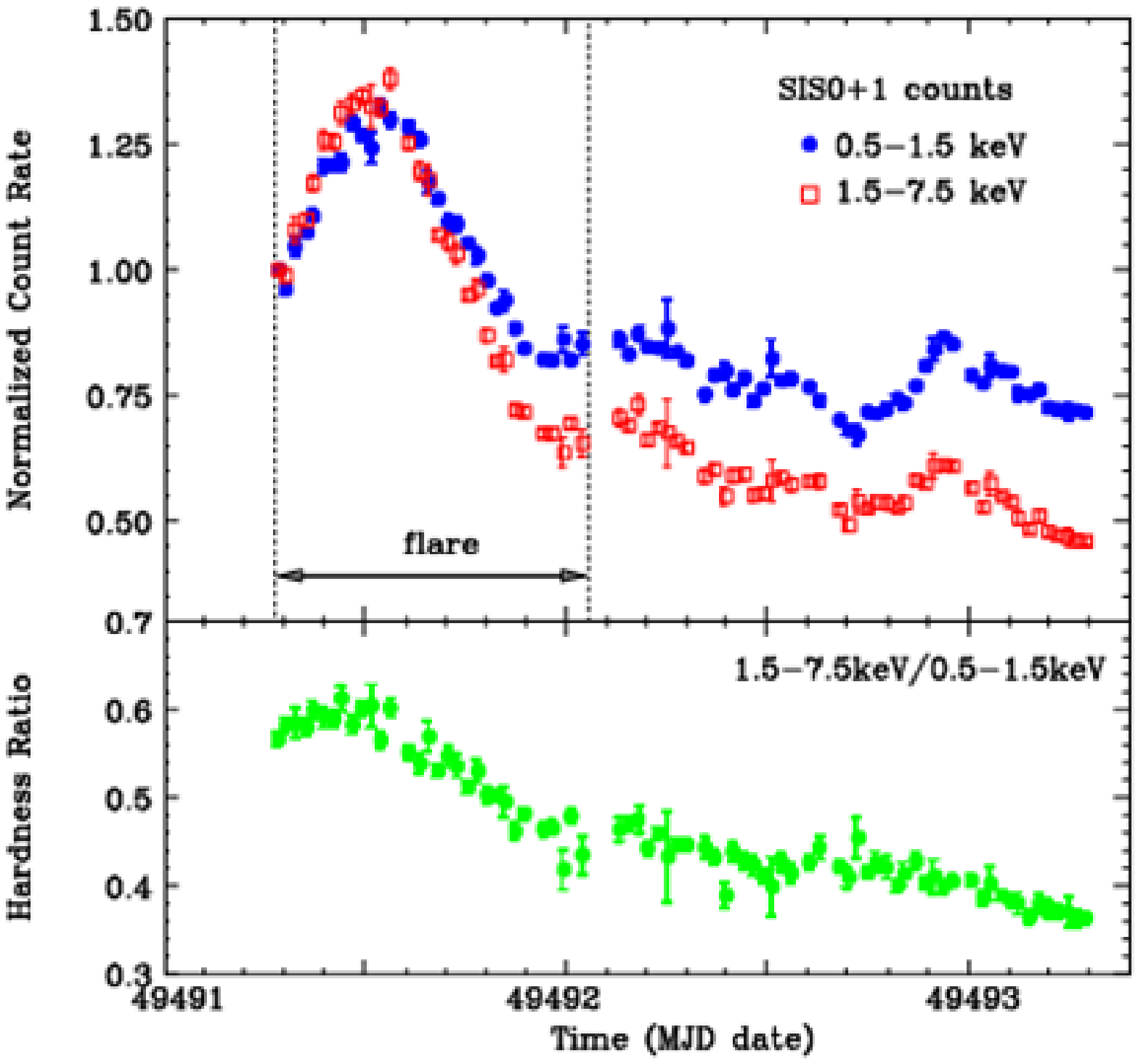}
  \includegraphics[height=.305\textheight]{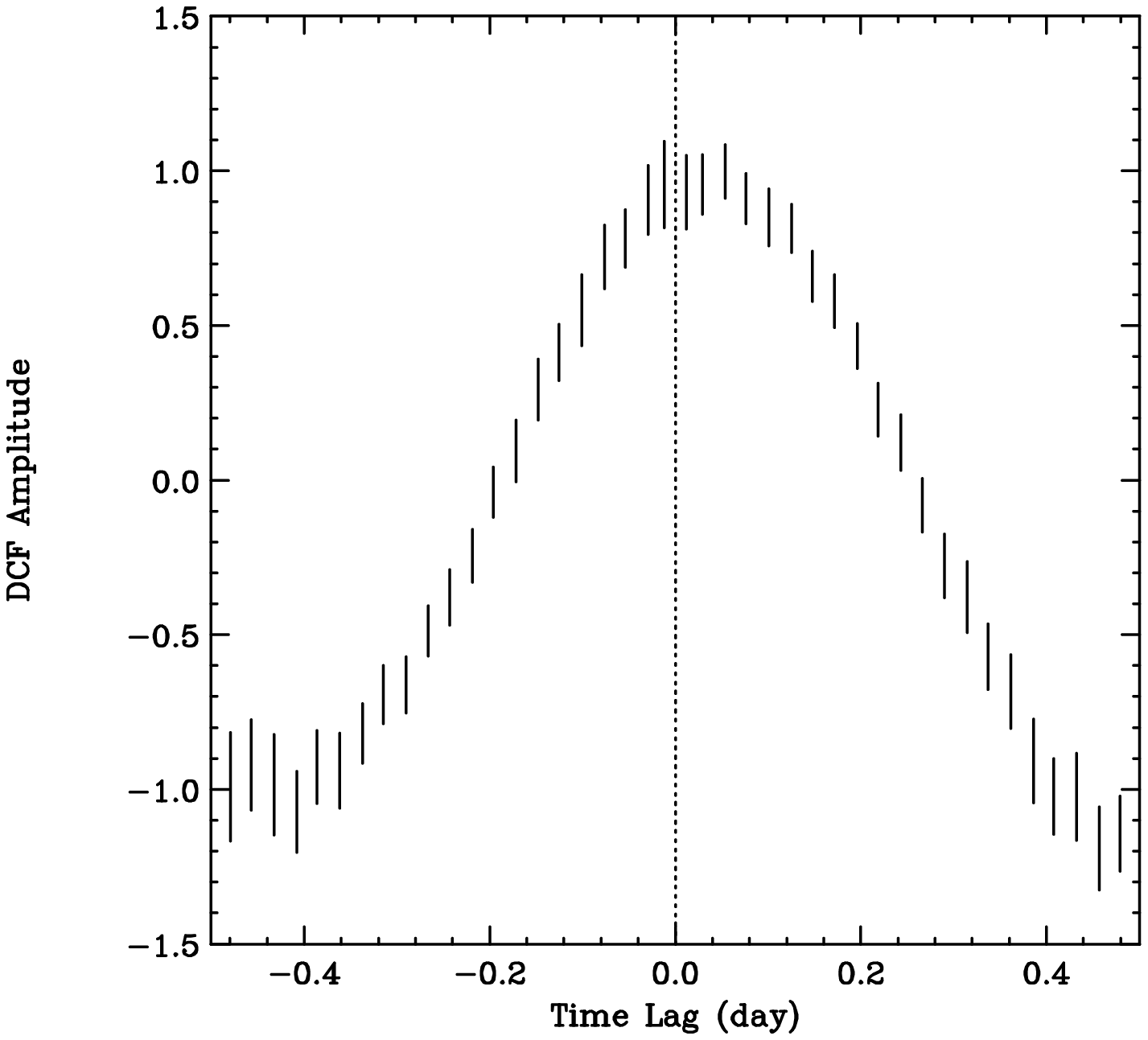}
  \caption{$left$: Time history of the X-ray emission of PKS 2155$-$304 
during 1994 observation with $ASCA$. Upper panel represents the X-ray 
right curves measured in 0.5$-$1.5 keV and 1.5$-$7.5 keV, respectively, 
while the bottom panel shows the time history of hardness ratio.
$right$: Discrete correlation function of PKS 2155$-$304 calculated from
the light curves in the $left$ panel. Time lags in 0.5$-$1.0 keV band 
was calculated as compared to that in the 3.0$-$7.5 keV band.}
\end{figure}

As an application, Figure~5 shows an example to see 
time-lag in the light curve of PKS 2155$-$304 (Kataoka et al. 2000). 
The data reveal a large flare at the beginning, followed by 
lower amplitude fluctuations. The source variability is 
somewhat different in different energy bands. Notably, 
amplitude of flux change is larger at higher photon energies; 
 a factor of 2 at 1.5$-$7.5 keV (red), while it is a factor 
of 1.5 in the 0.5$-$1.5 keV (blue). Also note that the peak of the light 
curve in the hard X-ray bands leads that in the soft X-ray bands 
by $\sim$ 4$-$5 ksec. This was also suggested by direct fitting 
of the light curves with a simple Gaussian plus constant offset, 
resulting that a lag of the peaking time by $\simeq$ 4 ksec.  
We therefore computed the cross correlations using the the DCF 
by dividing the 0.5$-$7.5 keV range into five energy bands and 
measured the time lag for each light curve compared to the 3.0
$-$7.5 keV light curve. The results are shown in Figure 5 ($right$), 
again suggesting $\simeq$ 4 ksec lags in the X-ray variability 
of PKS 2155$-$304.  

As we have seen in $\S$1, reality of this small amount of 
lag is still matter of debate, due to the periodic 
gaps ($\sim$ 6 ks) of low-Earth orbit satellites 
(e.g., Edelson et al. 2001). Meanwhile, it is also suggested that 
lags on hour-scale can hardly be produced by periodic gaps 
based on careful simulations (e.g., Tanihata et al. 2001; 
Zhang et al. 2004).  To quickly follow their arguments, 
I have made hundreds pairs of light curves by Monte Carlo simulation, 
one of which is artificially ``lagged'' by 4 ksec. Then the resultant 
light curves are filtered by the same window as the actual observation. 
Figure 6 ($left$) shows an example pair of light curves thus produced, 
and Figure 6 ($right$) shows the calculated DCF for 10 pairs of 
light curves. It seems that the DCF exhibits large uncertainties 
but the peak of the DCF is always retained as expected  (i.e., 4 ksec). 
Obviously, higher quality X-ray data which are less affected by 
window sampling is strongly awaited for further clarification 
of this long standing problem. 
At this point, it is also worth noting the most recent observation of a 
TeV blazar 1ES 1218+304 exhibiting a clear signature of 
time-lag which is much larger than the orbital gap ($\sim$ 20 ksec), 
but in the opposite sense (so-called ``hard-lag''). For more detail, see 
Sato et al. 2008 in this volume.
   
\begin{figure}
  \includegraphics[height=.285\textheight]{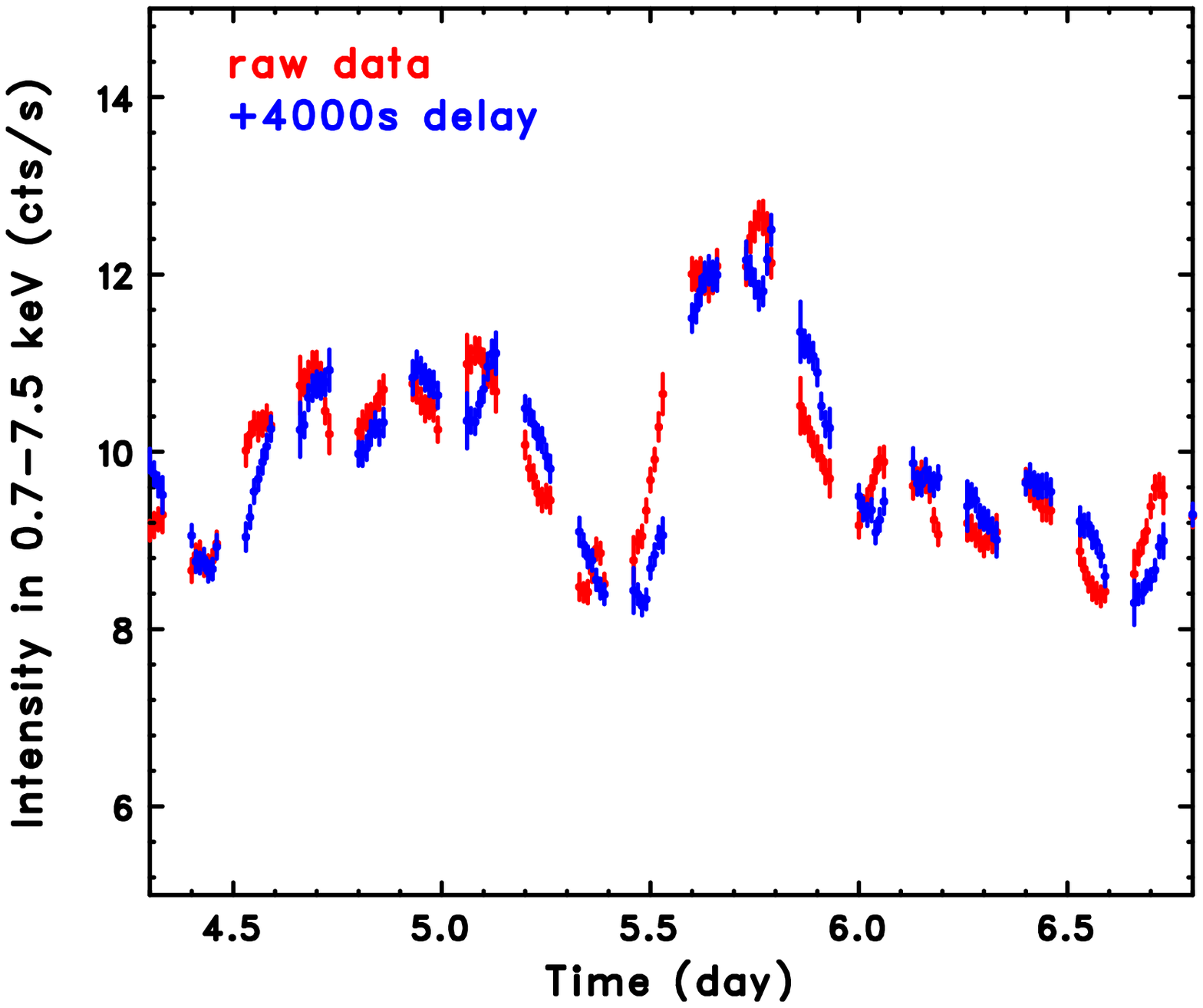}
  \includegraphics[height=.285\textheight]{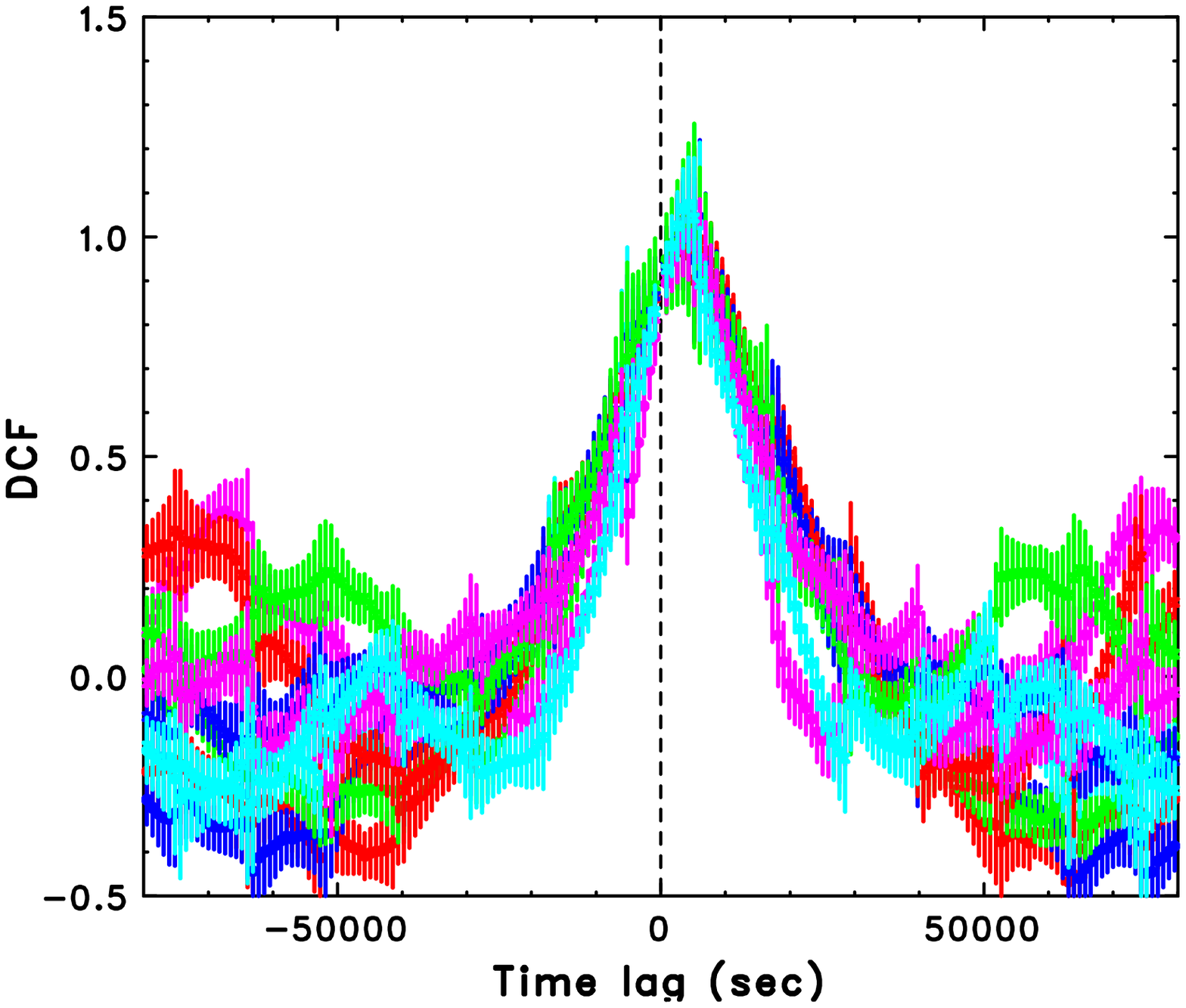}
  \caption{$left$: An example of simulated light curve to evaluate 
the effects caused by orbital gap in the measurement of time lag. 
$right$:  Examples of DCFs calculated from artificially lagged (by 4000
 sec) light  curves.}
\end{figure}

\section{Future Prospects; $GLAST$ and $MAXI$}

In the previous sections, we showed that temporal techniques, such as 
PSD, SF and DCF are indeed powerful tools to understand the nature of 
variability in blazars, as long as various artifacts are correctly 
taken into account. We showed that 
the Monte Calro simulation is the one of the best way to evaluate 
possible artifacts caused by window sampling as well as finite length 
of data. These approaches, however, are very time-consuming works 
and somewhat conservative for future progress. Very fortunately, 
we will have two important missions to uniformly/densely observe 
celestial sources in the high energy regime: 
$GLAST$ and $MAXI$. In particular, $GLAST$ was successfully launched 
June 11, 2008, and the activation of the LAT (Large Area Telescope) 
is about to begin (Madejski et al. 2008 in this volume).

\begin{table}
\begin{center}
\begin{tabular}{lrrrr}
\hline 
Source Name & Redshift & Class & Flux (2-10 keV)
& Flux ($\ge$100 MeV)\\
& &  & [10$^{-12}$ erg/cm$^2$/s] & [10$^{-5}$ ph/cm$^2$/s] \\

\hline
PKS 0208-512 & 1.00 & HPQ & 9.5    & 85.5$\pm$4.5\\
Q 0827+243   & 0.94 & LPQ & 4.8    & 24.9$\pm$3.9\\
PKS 1127-145 & 1.18 & LPQ & 11.0   & 38.3$\pm$8.0\\
PKS 1510-089 & 0.36 & LPQ & 10.0   & 18.0$\pm$3.8\\
3C~454.3 & 0.86 & HPQ & 11.0   & 53.7$\pm$4.0\\
3C~279 & 0.54 & HPQ & 13.0   & 89.0$\pm$3.2\\
PKS 0528+134 & 2.06 & LPQ & 30.0   & 60.0$\pm$3.0\\
\hline
\end{tabular}
\caption{A list of ``VIP'' blazars to be simultaneously observed with 
$GLAST$ and $Suzaku$ in 2008/09.}
\label{tab:a}
\end{center}
\end{table}

It is widely expected that $GLAST$ will detect a 
large number (probably between 3,000 and 10,000) of 
extragalactic sources, most of which will be identified as blazars. 
Moreover, the LAT large field-of-view combined with scanning 
mode will provide a  very uniform exposure over the sky, 
allowing constant monitoring of all detected blazars and 
flare alerts to be issued.  Apparently,  simultaneous 
multiwavelength campaigns are essentially important for both ``EGRET blazars'' 
(i.e., well-established sources) as well as newly detected $\gamma$-ray 
sources. In X-ray, many observatories are already being actively prepared. 
For example, we are planning dedicated campaigns of 7 quasar hosted 
blazars (QHBs) as a part  
of $Suzaku$-AO3 as listed in Table 1. Assuming a large flare 
as that observed for 3C~279 in 1991, $Suzaku$ can  determine the 
X-ray spectrum up to 300 keV with an unprecedented accuracy. 
Coordinated observations between $GLAST$ and X-ray satellites  
are crucial for further understanding the nature of various types 
of blazars.

\begin{figure}
\begin{center}
  \includegraphics[height=.33\textheight]{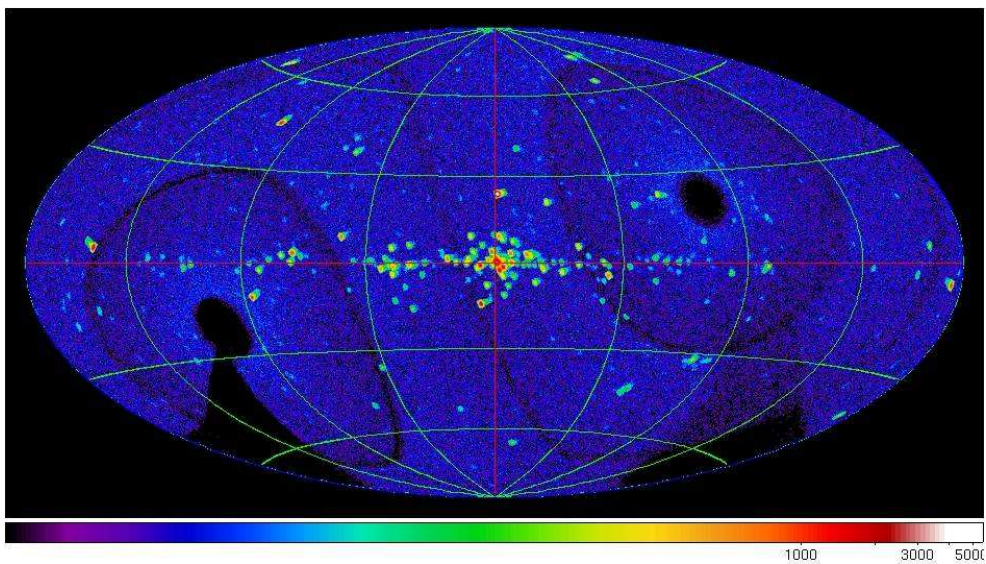}
  \caption{An expected X-ray sky map for 1 day exposure with $MAXI$.}
\end{center}
\end{figure}

\begin{figure}
  \includegraphics[height=.283\textheight]{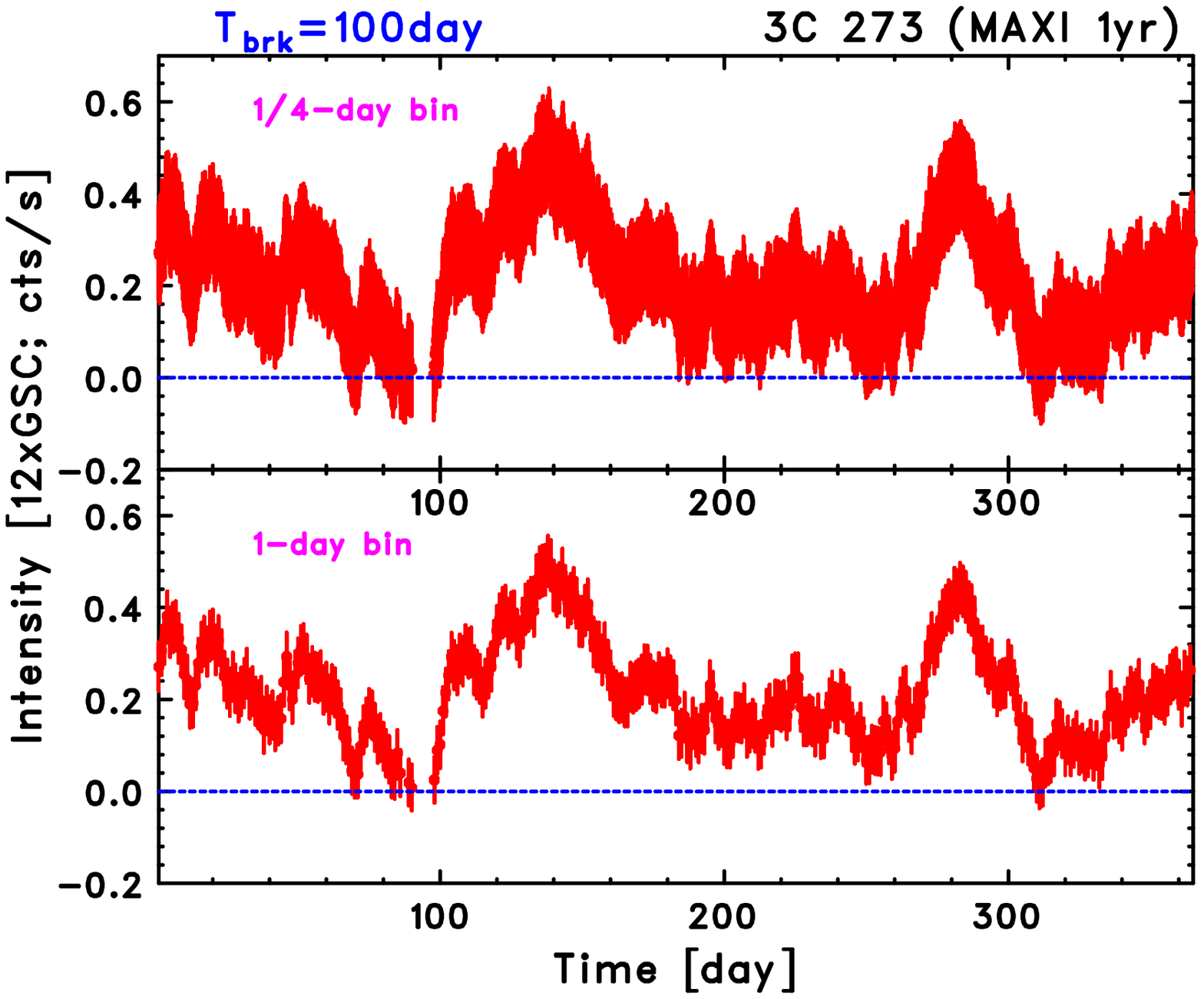}
  \includegraphics[height=.283\textheight]{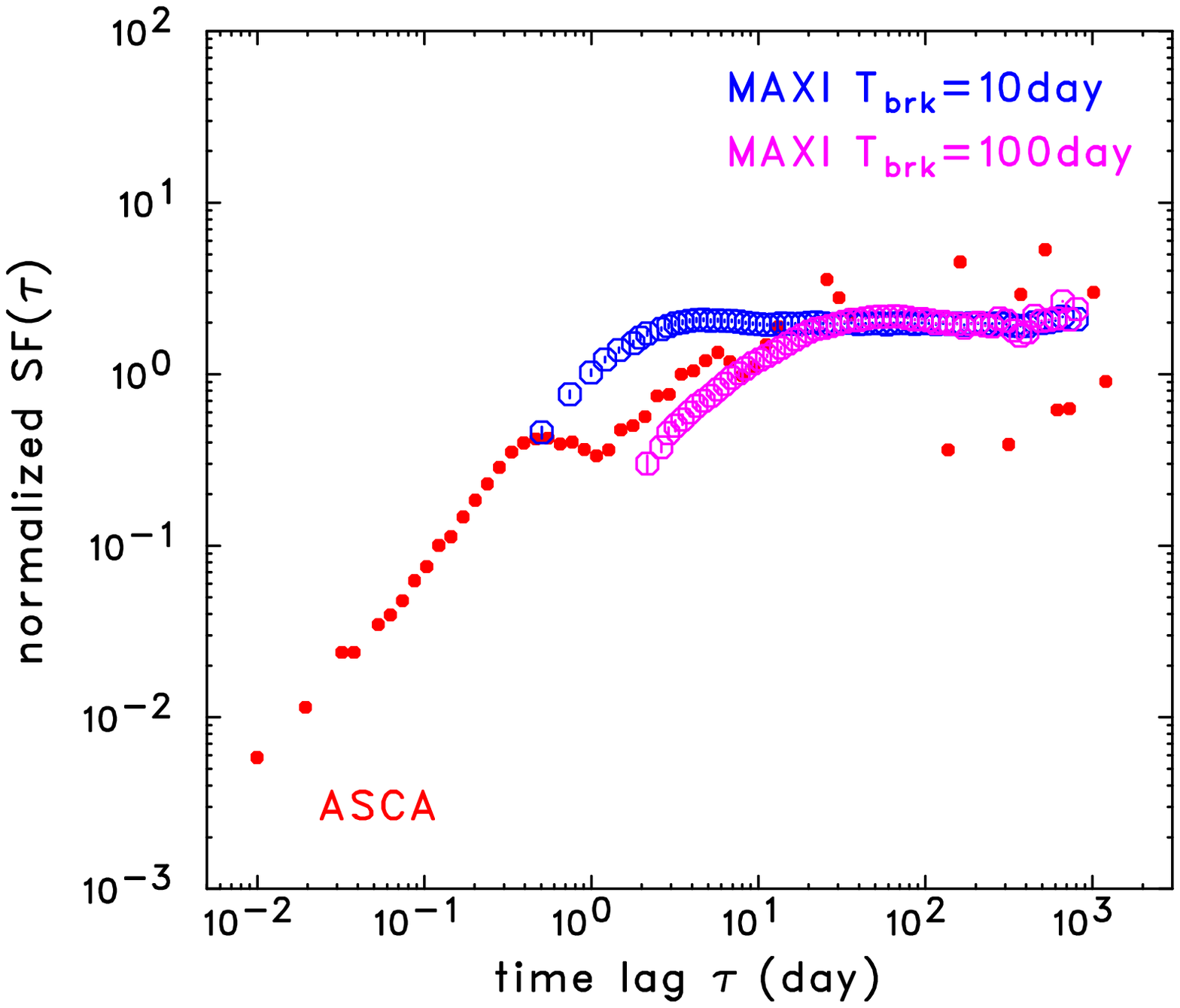}
  \caption{$left$: Simulation of 3C~273 light curve observed 
with $MAXI$. Data are binned at 0.25 day ($upper$) 
and 1 day ($lower$), respectively. $right$: Structure function of Mrk 421 
calculated from the simulated light curves with $MAXI$ as a 10 mCrab 
source.} 
\end{figure}

Another important mission for future blazar studies will be 
the Monitor of All-sky X-ray Image ($MAXI$). $MAXI$ is an X-ray all-sky 
monitor which is currently scheduled to be attached to the Japanese
Experiment Module - Exposed Facility (JEM-EF) on the  International
Space Station (ISS) in early 2009. The $MAXI$ carries two scientific 
instruments: the Gas Slit Camera (GSC) and the Solid 
State-slit Camera (SSC). The GSC consists of position-sensitive 
proportional counters with large collecting area of 5350 cm$^2$ in 
2$-$30 keV range, while the SSC is utilizing 32 X-ray CCD chips
covering an energy range of 0.5-12 keV. The $MAXI$ has two sets of 
GSC and SSC orthogonally oriented, each of which covers a narrow 
instantaneous field of view of 1.5 deg times 160 deg that 
sweeps over the whole sky during every orbit of 90 minutes. 
Thus a certain sky area is generally monitored twice in an orbit.
The expected detection sensitivity for the GSC is $\sim$5 mCrab 
in a day and $\sim$1 mCrab in one month, which is higher by a factor 
of 5 than that of $RXTE$/ASM. Such a high sensitivity and its monitor 
capability are very useful in study of AGNs, compact sources 
such as microquasars and Galactic blackholes.

Figure 7 shows the expected X-ray sky map for 1 day exposure with $MAXI$.
Bright AGNs, such as 3C~273 ($\simeq$ 5 mCrab) and Mrk 421 ($\simeq$ 10 
mCrab) can be detected with more than 5 $\sigma$ level everyday,
allowing for the first time non-bias monitoring of the sources from 
a day to more than year scale.  Figure 8 ($left$) shows the simulated 
long-term (1 year) light curve of 3C~273, 
assuming a PSD slope of 2.0 with break 
time scale of $1/f_{\rm brk}$ $\simeq$ 100 day. The resultant 
structure function  of the light curve, presented as Figure 8 ($right$),
clearly revels variability 
nature of blazars on extremely longer timescale than the
characteristic break. It is suggested that these long-term trend may be 
produced by the time variation of accreting matter near the central
black hole, as well as the duty cycle of mass ejection to the
relativistic jet, which provide important challenges to blazars 
in the next decade.

\end{document}